\documentstyle[12pt]{article} 
 
\begin{document}
\def\la{\mathrel{\mathpalette\fun <}}
\def\ga{\mathrel{\mathpalette\fun >}}
\def\fun#1#2{\lower3.6pt\vbox{\baselineskip0pt\lineskip.9pt
        \ialign{$\mathsurround=0pt#1\hfill##\hfil$\crcr#2\crcr\sim\crcr}}}
\newcommand {\eegg}{e^+e^-\gamma\gamma~+\not \! \!{E}_T}
\newcommand {\mumugg}{\mu^+\mu^-\gamma\gamma~+\not \! \!{E}_T}
\renewcommand{\thefootnote}{\fnsymbol{footnote}}
\bibliographystyle{unsrt}

\begin{flushright}
UMD-PP-99-053\\
\end{flushright} 
\begin{center}
{\Large \bf Diffuse Ionization in the Milky Way and Sterile Neutrinos}

\vskip1.0cm

{\bf Rabindra N. Mohapatra$^1$, and Dennis W. Sciama$^2$}

\vskip1.0cm

{\it$^{(1)}${ Department of Physics, University of Maryland,
College Park, MD-20742, USA.}}

{\it $^{(2)}${SISSA and ICTP, Trieste and Dept. of Physics, University of
Oxford, UK}}

\end{center}

\begin{abstract} 
We propose the radiative decay of sterile neutrinos which fill a fraction of 
the halo dark matter with a mass of 27.4 eV and lifetime of $\sim 10^{22}$ 
sec. as a way to
explain the observed diffuse ionization in the Milky Way galaxy. Since 
the sterile neutrino number density in the present universe can be 
adjusted by arranging its dynamics appropriately, the resulting hot dark 
matter contribution to $\Omega_m$ can be small as required by many scenarios of 
structure formation. On the other hand a 27.4 eV neutrino could easily be 
the partial halo dark matter. One realization of this idea could be in the 
context of a mirror universe theory where the gauge and matter content of 
the standard model are completely duplicated in the mirror sector 
(except for an  asymmetry in the weak scale); the three mirror neutrinos 
can mix with 
the known neutrinos via some strongly suppressed mechanism such as the 
gravitational or heavy right handed neutrino mediated forces. 
Two of the mirror neutrinos (say $\nu'_{\mu}$ and $\nu'_{\tau}$) 
could play the role of the above sterile neutrino.

\end{abstract}

\newpage
\renewcommand{\thefootnote}{\arabic{footnote})}
\setcounter{footnote}{0}
\addtocounter{page}{-1}
\baselineskip=24pt

 \vskip0.5cm

The interstellar medium of the Milky Way galaxy is known to contain
ionized hydrogen gas with properties which seem to defy common 
astrophysical explanations\cite{sciama1}. Similar problems also seem to 
exist for other galaxies studied. One solution that proposes to 
use a decaying dark matter neutrino was suggested by Sciama\cite{sciama1}. 
According to this idea, if the relic tau neutrinos which are supposed to 
pervade the cosmic background with a density of $\approx 113 (cm)^{-3}$ 
have a mass of 27.4 eV and 
decay to one of the other two lighter neutrinos plus a photon with a 
lifetime of about $2\times 10^{23}$ sec. then they could provide enough 
ionizing photons on 
galaxy scale heights of nearly a kilopersec to solve this problem. The 
reason for this is that the decay photon has an energy of 13.7 eV and
so would be able to ionize hydrogen-the lifetime is chosen to ensure
that the resulting photon flux would produce the observed electron density.
With these parameters the tau neutrino could be the dark matter needed 
to account for the flat rotation curve of the galaxy. Despite the 
attractiveness of this suggestion, the 
fact that a 27.4 eV neutrino must provide 100\% of the dark matter of the 
universe runs into trouble with the standard scenarios for structure 
formation. Also the fact that this scenario implies that the $\Omega_m=1$
may be in conflict with  high z type Ia supernova data\cite{perl}  as 
well as data from clusters\cite{eke}, which suggest that $\Omega_m$ may be 
considerably less than 
unity. Although one could use cosmic strings to get out of the first
difficulty, it is tempting to 
look for alternative proposals that may keep the basic ingredients of the 
idea and yet not force us into a 100\% HDM-full universe.

In order to set the stage for our suggestion, let us note that the reason 
why an active neutrino of mass 27.4 eV constitutes nearly 100\% of the 
dark matter in the universe
is that its normal weak interaction allows it to remain in equilibrium 
until the universe is a second old. After it decouples, the annihilation 
of $e^+e^-$ to photons increases the temperature of the universe 
slightly leaving the neutrinos unaffected. Thus at the present epoch 
their number density becomes somewhat lower than that of the photons 
yielding about 113 neutrinos of each type per cm$^3$. For the Hubble 
parameter $h$ of about $0.5$, this leads to $\Omega_{\nu}=1$. On the other 
hand if we had a sterile neutrino of mass 27.4 eV, its present number 
density will be 
controlled by its interactions with the standard model particles which 
by virtue of its being sterile are much weaker. This makes  it decouple 
earlier from the cosmic soup (say at $T\geq 200$ MeV), thus making its 
number density much smaller (since the number of degrees of freedom 
contributing to the energy density of the universe changes drastically 
below the QCD phase transition temperature which happens to be around 200 
MeV). Their contributions to $\Omega$ is therefore much smaller.
These relic sterile neutrinos will be uniformly distributed thoughout the
universe and will eventually concentrate into the halos of galaxies after 
they form constituting a fraction of the halo dark matter (which could be 
about 10\% or so). If its lifetime to photonic decay is proportionally reduced,
then it could explain the problem of diffuse ionization in our Galaxy 
while at the same time avoiding being the dominant dark matter constituent.
In this letter we elaborate on this proposal, seek gauge theories where
this proposal can be realized and suggest some tests.

\begin{center}
{\bf I. The sterile neutrino mass and lifetime implied by the solution to the
diffuse ionization problem}
\end{center}

To introduce the requirements for the sterile neutrino to 
solve the diffuse ionization problem, let us recapitulate some of the basic 
issues involved\cite{sciama1}. The major problem that needs a 
solution is the observed diffuse ionization in the Milky Way,
specifically its scale height of nearly 670 pc and its rather
uniform distribution in different directions. 
Suggestions involving the cosmic rays 
have been ruled out by observations\cite{mes} and any local sources for 
hydrogen ionizing radiation must overcome the problem of opacity due to 
neutral hydrogen HI, that makes it difficult to understand the magnitude 
of the scale height. On the other hand if an all pervading medium of relic
neutrinos decay to photon with an energy of 13.7 eV, it easily evade all 
these problems. 

The key formula for our purpose is the one that dictates the 
equilibrium between the process of hydrogen ionization by the photons from 
neutrino decay and recombination of free electron with proton to form 
neutral hydrogen back. If $\alpha$ is assumed to denote the recombination 
rate of $e^- +p\rightarrow H$, the equilibrium condition is given by:
\begin{eqnarray}
\frac{n_\nu}{\tau_\nu}=\alpha n^2_e
\end{eqnarray}
For the relevant electron temperature of $10^4$ K, the value of $\alpha 
\simeq 2.6\times 10^{-13}$ cm$^3$ sec$^{-1}$. Taking $n_e\sim 0.033$ 
cm$^{-3}$ known from observations, we find that 
$\frac{n_{\nu}}{\tau_{\nu}}\sim 2\times 10^{-16}$
cm$^{-3}$ sec$^{-1}$. If we now assume that the neutrino 
lifetime is $2\times 10^{23}$ sec. this is then consistent with the 
various constraints from supernova 1987A observations\cite{kolb}. This 
requires the halo density of neutrinos to be about
$4\times 10^{7}$ cm$^{-3}$. As has been shown by Sciama, within a simple 
approximation of isothermal and isotropic distribution of halo dark matter 
\cite{ost} of neutrinos, the above number density emerges quite naturally.
A simple way to see this is to note that the halo density is known from 
fits to the galaxy rotation curves to be about $\rho \sim .3$ GeV cm$^{-3}$.
The formula $\rho\sim n_{\nu} m_{\nu}$ then yields the above number for 
the halo density of tau neutrinos.

This discussion can be translated to the case of sterile neutrinos, for
whom the relic density will be given by:
\begin{eqnarray}
\frac{n_{\nu'}}{n_{\gamma}}\simeq \frac{g_{*}(T_0)}{g_{*}(200 MeV)}
\end{eqnarray}
where $g^*(T)$ is the number of degrees of freedom of the particle species
in equilibrium with electrons at the temperature, T.
If we assume that the QCD phase transition temperature is below 200 MeV, then
we get $g_*(200 MeV)= \frac{261}{4}$ and $g_*(T_0)=2$ where 
$T_0$ is the present temperature of the universe. This gives, $n_{\nu'}\simeq
12$. This leads to $\Omega_{\nu'}\simeq 0.08$. Thus the sterile neutrinos 
do not constitute a significant part of the dark matter "menu". If we 
further assume that $\Omega_m\simeq 0.4$ and $\Omega_B\simeq 0.08$ and the 
same ratio is maintained for all the particle species in the dark halo, the 
halo density of sterile neutrinos will be $\simeq 2\times 10^{6}$ cm$^{-3}$.
Eq. (1) then dictates that we must have a radiative lifetime of the 
sterile neutrino of $10^{22}$ sec. for it to be able to explain the 
diffuse ionization problem.

Before getting into the particle physics models let us first note how 
such a radiative decay lifetime can be achieved for the sterile neutrino.
Recall that in order for the sterile neutrino to be useful in resolving 
the neutrino puzzles as we will eventually assume, the active and the 
sterile neutrinos must mix with each other. But for this mixing not to effect
the big bang nucleosynthesis constraints\cite{subir}, the masses and mixings
must satisfy the constraint\cite{kimo}\footnote{Note however,  a
recent arguement by Foot and Volkas\cite{foot} according to which the 
bounds on sterile-active neutrino mixing could be considerably weaker 
for certain parameter ranges due to lepton asymmetry generated before 
the big bang nucleosynthesis epoch. This would help us in expanding the 
allowed parameter space of our proposal}:
 \begin{eqnarray}
\Delta m^2_{\nu_a\nu'} (sin^42\theta) \leq 2\times 10^{-6} eV^2
\end{eqnarray}
Since in our case, $\Delta m^2\simeq 10^3$ eV$^2$, we get, $sin 2\theta\leq
0.7\times 10^{-2}$. Now suppose that there is a transition magnetic moment
involving the $\nu_{\mu}-\nu_{\tau}$ ($\mu_{23}$). Then,
the radiative decay of the sterile neutrino can occur via its mixing with 
either the $\nu_{\mu}$ or $\nu_{\tau}$. This will lead to a decay rate 
\begin{eqnarray}
\Gamma_{\nu'\rightarrow \nu_{\mu}+\gamma}\simeq \frac{\theta^2 
\mu^2_{23}}{8\pi}m^3_{\nu'}
\end{eqnarray}
Using the afore mentioned values for the parameters in the above 
expression, we get
a lifetime for the sterile neutrino $\tau_{\nu'}\simeq 10^{22}$ sec. for 
the choice of $\mu_{23}\simeq  10^{-12}\mu_B$ for $\theta\simeq 
0.01$. Thus we get the desired radiative lifetime to solve the diffuse 
ionization problem. Since we have quite reasonable values for all the 
necessary parameters involved in our mechanism, we feel that this is a viable
solution to the problem at hand.

A point that needs to be noted here is that the supernova bound for the 
radiative lifetime for the eV neutrinos need not apply to our case since we 
are considering a sterile neutrino and not an active one. Its production in 
the supernova will be different from the active neutrino case.

This brings us to raise the two key particle physics issues that we must
address: is it possible to have a viable model for a sterile neutrino 
whose mass of 27.4 eV is not unnatural and how does it fit into the full
neutrino picture that accomodates the observations or indications of 
neutrino oscillations in solar, atmospheric as well as the LSND data.
We address these questions in the next section.

\begin{center}
{\bf II. The Complete Scenario for Neutrino Puzzles:}
\end{center}

There are various scenarios for neutrino masses that can be constructed 
to fit a radiatively decaying 27.4 eV sterile neutrino. For simplicity of
presentation, we focus on a variation of the following six neutrino picture 
that emerges in 
the context of the mirror universe model for particle physics\cite{moh}
and assume the following pattern for the neutrino masses: $m^2_{\nu'_e}-
m^2_{\nu_e}\simeq 10^{-5}$ eV$^2$; $m_{\nu_{\mu}}\simeq m_{\nu_{\tau}}\sim
\sqrt {{\Delta m^2}_{LSND}}\sim .2~eV-3~eV$ and $m_{\nu'_{\mu}}\simeq 
m_{\nu'_{\tau}}\sim 27.4$ eV. Except perhaps for a possible mirror 
symmetry, there is no reason for the $\nu'_{\mu}$ and $\nu'_{\tau}$ to 
have same mass and when needed we can revert to a scheme with only one of
$\nu'_{\mu,\tau}$ having a mass of 27.4 eV and the other much lighter.
Note that the primed neutrinos do not couple to the standard model gauge 
group and are therefore the sterile neutrinos.

In this picture, the solar neutrino puzzle is solved via $\nu_e-\nu'_e$ 
oscillation whereas, the atmospheric neutrino oscillation is between
the $\nu_{\mu}-\nu_{\tau}$ as suggested in Ref.\cite{cald}. The 27.4 eV
$\nu'_{\mu,\tau}$ will play the role in solving the diffuse ionization 
problem. Next, we present a gauge model where the 
radiative decay with the required lifetime can emerge. The basic idea is that
we must generate a reasonable transition magnetic moment between the active
$\nu_{\mu}$ and $\nu_{\tau}$. For this we can choose a version of MSSM with
R-parity violating interactions as in Ref.\cite{babu}. Whereas we have 
chosen this particular model for the purpose of illustration, one could 
use any other model that generates a large neutrino magnetic moment while 
at the same time keeping the mass small\cite{all}.

Let us consider the gauge group of the model to be $G\otimes G'$ where
G and G' are identical groups with G operating on the visible sector 
fields which contain the standard model particles and G' operating on the 
mirror sector fields which have an identical spectrum as the visible sector.
We choose $G\equiv G'= SU(3)_c\otimes SU(2)_L\otimes U(1)_{Y}$. 
(All groups in the mirror sector will be denoted by a primed 
symbol). The spectrum of matter and Higgs fields for each sector is same
as in the MSSM. The mirror sector fields will be denoted by a prime on the 
above fields. The basic idea of the class of models we will be interested in
is that they will generate a transition magnetic moment $\mu_{23}\sim 
10^{-12}\mu_B$ while at the same time keeping all neutrino masses $\leq$1 eV.

For the superpotential, we choose,
\begin{eqnarray}
W~=~h_uQH_uu^c + h_d Q H_d d^c + L H_d e^c+ MH_uH_d \nonumber\\+ 
+f(L_{\mu}L_e\mu^c + L_{\tau}L_e\tau^c) + f_1 L_{\mu}L_{\tau} e^c +L_eH_de^c
\nonumber\\
h_{\mu}(L_{\mu}H_d\mu^c +L_{\tau}H_d\tau^c) 
\end{eqnarray}
Note that this superpotential has an $SU(2)_H$ symmetry between the 
$L_{\mu}$ and $L_{\tau}$ (as between ($\mu^c, \tau^c$)). We break this 
symmetry softly in the supersymmetry breaking sector
by the slepton mass terms being different. As in Ref.\cite{babu}, this 
model gives rise to a large transition magnetic moment for the $\mu$ and 
$\tau$ neutrinos without simultaneously giving them large masses. 
The value of $\mu_{23}$ is given by 
\begin{eqnarray}
\mu_{23}\simeq \frac{ff_1e}{8\pi^2} m_{\tau} sin 2\phi 
\left(\frac{1}{M^2}ln\frac{M^2}{m^2_e}\right)\simeq ff_1 5\times 10^{-8}\mu_B
\end{eqnarray}
where $\phi$ denotes a mixing angle in the Higgs sector and $M$ a typical 
Higgs mass for which we have chosen a value of 100 GeV. Thus to get the
desired value for the magnetic moment, we need $ff_1\simeq 2\times 10^{-5}$.
Note that in the absence of 
symmetry breaking the muon and the tau have the same mass. But it is well 
known that supersymmetry breaking terms can be used to generate this 
splitting if they are appropriately chosen. We do not go into this 
discussion since this is identical to what is in Ref.\cite{babu}. One needs
a certain degree of fine tuning to achieve the desired values. However,
since our goal is to give plausible arguments for the kind of parameters 
we use in our proposal, we refrain from getting into the full naturalness 
discussion.

Turning now to the discussion of the muon and tau neutrino masses, we see 
that in this model, the dominant contribution comes from the radiative 
one loop diagram and leads to an off diagonal mass matrix in the lowest 
order. Again in the symmetry limit, they vanish. Their magnitude can be 
calculated to be: \begin{eqnarray}
m_{\nu_{\mu}-\nu_{\tau}}\simeq \frac{ff_1}{16\pi^2} m_{\tau} F(M^2, {M'}^2, 
\delta,\delta')
\end{eqnarray}
where $M, M', \delta,\delta'$ are parameters characterising the 
supersymmetry breaking sector of the second and the third generation.
It is not impossible to arrange these parameters to get this mixing mass 
to be in the few eV range, since all one needs to do is to have $F\approx 
0.1$. The key point here is that in the limit of the 
exact $SU(2)_H$ symmetry, $F=0$ whereas $\mu_{23}\neq 0$. This property 
enables us to maintain a degree of naturalness in generating the small 
neutrino masses. Note that this mass matrix gives rise to a 
maximal mixing pattern for the $\nu_{\mu}-\nu_{\tau}$ sector as required 
by the atmospheric neutrino data. The mass splitting between them must 
come from alternative sources. One way for example is to put in a doubly 
charged field $\Delta \oplus \bar{\Delta}$ which is an $SU(2)_L$ singlet
with a coupling in the superpotential $\Delta e^ce^c$ and a symmetry 
violating term in the soft
breaking term of type $\tilde{m}\tilde{\mu^c}\tilde{\mu^c}\tilde{\Delta}$. 
This generates a neutrino mass
of type $\nu_{\mu}\nu_{\mu}$ at the two loop level. Its value can be 
estimated to be of order
\begin{eqnarray}
m_{\nu_{\mu}\nu_{\mu}} \simeq \frac{f^2}{(16\pi^2)^2} \frac{m^2_e}{M}
\end{eqnarray}  so that
the above diagonal term is of order $0.03$ eV which is in the right 
range for the atmospheric neutrino data.

This model leads to a one loop contribution to the $\nu_e$ mass given by
\begin{eqnarray}
m_{\nu_e}\simeq \frac{f^2}{16\pi^2} 
m_{\tau}ln\frac{m^2_{\tau}}{m^2_{\tilde{l}}} \end{eqnarray} 
which is easily in the eV range if we require that $f\sim 10^{-4}$. Thus the 
understanding of the solar neutrino problem would require very fine tuned 
vacuum oscillation between the $\nu_e-\nu'_e$. This would require that the 
soft slepton masses must be asymmetric between the normal and the mirror 
sector so that $m_{\nu_e}\simeq m_{\nu'_e}$.

The mirror sector of the model in the limit of exact gauge symmetry and 
supersymmetry is assumed to 
be a complete duplicate of the visible sector. After electroweak symmetry 
breaking, we will choose the symmetry breaking vev $v'_{wk}$ to be much 
larger (about a factor of 30) than the $v_{wk}$. Similarly the slepton 
masses 
which result from supersymmetry breaking will also be required to be 
different in the two sectors.
 The mixing between the normal and mirror sector arise via the higher 
dimensional operators such as $L_{\mu}H_uL_{\mu'}H'_u/M_{Pl}$ etc as in 
Ref.\cite{moh} and as demonstrated there, one can have eV range masses 
for the $\nu_{\mu,\tau}$ and 30 eV mass for the $\nu'_{\tau}$, enabling 
our mechanism to operate.

Coming to the $\nu_e$ sector, similar arguments also imply a mixing mass
of order $10^{-3}$ eV and if $m_{\nu_e}\simeq 0.1$ eV, this gives
$sin^22\theta_{\nu_e-\nu'_e}\simeq 10^{-2}$ which is in the right range for 
the MSW solar neutrino oscillation via the small angle oscillation.

We thus see that it is possible to construct plausible particle physics 
models for our proposal.

It is worth pointing out that the large transition magnetic moment 
between $\nu_{\mu}$ and $\nu_{tau}$ which is an essential ingredient
of our proposal has other applications. It has been suggested\cite{lanza}
that it can provide a mechanism to revive the stalled shock in supernovae.
Another application is its usefulness in understanding pulsar velocities
using resonant spin flavor transition\cite{akh} in the neutrino 
sphere of supernovae\footnote{ As has been noted by Kusenko and 
Segre\cite{segre}, one
could also explain pulsar velocities using massive sterile neutrinos.}.

In conclusion, we have presented a modified version of the original
tau neutrino radiative decay model for understanding the diffuse 
ionization in the Milky Way and other galaxies. The model avoids 
difficulties with structure formation. One particle physics realization 
of the idea is presented in the context of the already proposed mirror 
universe idea for neutrino puzzles. 

The work of R. N. M. is supported by the National Science 
Foundation grant No. PHY-9802551.

\end{document}